\documentclass[prl,amsmath,amssymb,twocolumn]{revtex4}
\usepackage{mathrsfs}
\usepackage{amssymb}

\input epsf.tex
\usepackage{graphicx}
\usepackage{amsthm}

\begin{document}

\title{Attacking practical quantum key distribution system with wavelength dependent beam splitter and multi-wavelength sources}

\author{Hong-Wei Li$^{1,2,*}$, Shuang Wang$^{1,*}$, Jing-Zheng Huang$^{1,*}$,  Wei Chen$^{1,a}$, Zhen-Qiang Yin$^{1,b}$, Fang-Yi
Li$^{1}$, Zheng Zhou$^{1}$, Dong Liu$^{1}$, Yang Zhang$^{1}$,
Guang-Can Guo$^1$, Wan-Su Bao$^{2}$, Zheng-Fu Han$^{1c}$ }

 \affiliation
 {$^1$ Key Laboratory of Quantum Information,University of Science and Technology of China,Hefei, 230026,
 China\\$^2$ Zhengzhou Information Science and Technology Institute, Zhengzhou, 450004,
 China}

 \date{\today}
\begin{abstract}
It is well known that unconditional security of quantum key
distribution(QKD) can be guaranteed by quantum mechanics. However,
the practical QKD systems have some imperfections, which can be
controlled by the eavesdropper to attack the secret key. With the
current experimental technology, the realistic beam splitter, made
by the fused biconical technology, has wavelength dependent
property. Based on this fatal security loophole, we propose
wavelength-dependent attacking protocol, which can be applied to all
practical QKD systems with the passive state modulation. Moreover,
we experimentally attack practical polarization encoding QKD system
to get all the secret key information at the cost of only increasing
quantum bit error rate from $1.3\%$ to $1.4\%$.
\end{abstract}
\maketitle

Quantum key distribution is the art of sharing secret keys between
two distant parties Alice and Bob. Since the BB84 protocol has been
proposed by Bennet and Brassard \cite{BB84}, the unconditional
security of QKD protocol has attracted much attentions. Lo and Chau
\cite{Lo} proved unconditional security of BB84 protocol with
quantum computer. Shor and Preskill \cite{Shor} proved unconditional
security of BB84 protocol by applying the entanglement distillation
and purification (EDP) technology. More recently, Renner
\cite{Renner} proved unconditional security of BB84 protocol by
applying the quantum information theory method.

Whereas, security analysis model based on the perfect QKD protocol
can not be directly applied to the practical QKD systems
\cite{Shuang,Yamamoto,Stucki}, Gottesman, Lo, Lukenhaus and Preskill
\cite{GLLP} analyzed security of the practical QKD system and gave
the famous secret key rate formula GLLP. Combining their security
analysis result with decoy state method \cite{decoy theory1, decoy
theory2, decoy theory3}, practical QKD system can be realized with
weak coherent source. But their security analysis can not be applied
to the practical QKD system with arbitrary imperfections \cite{Li1,
Li2}, which may introduce side channel attacks. Imperfect phase
modulator introducing phase-remapping attack has been experimentally
demonstrated \cite{Xu}. Imperfect single photon detector (SPD)
introducing detector blinding attack has also been proposed in Ref.
\cite{Lars}, they demonstrated that imperfect SPD can be fully
remote-controlled by utilizing specially tailored bright
illumination. More recently, dead time attack with imperfect SPD has
been proposed in Ref. \cite{Henning}, in which the eavesdropper can
exploit the dead time effect of the imperfect SPD to gain almost
full secret information without being discovered. Jain et al.
\cite{Jain} have proved that inappropriately implemented calibration
routine will introduce a fatal security loophole. All these results
demonstrate that practical QKD device imperfections can lead to
various types of attacks \cite{HA0,HA1,HA2,HA3,HA4,HA5}. In current
experimental realizations, beam splitter has the wavelength
dependent property. Based on this imperfection, we propose a new
type of attacking protocol. Our experimental demonstration shows
that this strategy can effectively attack practical passive
modulated polarization based QKD system without being discovered,
where passive(active) modulation implies that Bob
passively(actively) select measurement bases. It should be noted
that the attacking model can also be easily generalized to other
passive modulated QKD systems.

Practical QKD systems can be divided into phase encoded and
polarization encoded respectively. In the polarization based QKD
systems \cite{PB1,PB2}, Bob passively selects the measurement basis
by the BS for convenient and high speed modulation. More precisely,
the $1\times2$ BS has one input port and two output ports (port 1
and port 2), Bob can choose to measure the photon state either in
rectilinear basis if it pass through output port 1, or in diagonal
basis if it pass through output port 2. In the perfect case, the
single photon state will randomly select to pass through one output
port of the BS. But, the realistic BS is commonly made by the fused
biconical taper (FBT) technology \cite{BS0}, the coupling ratio of
the FBT BS is generally wavelength-dependent. We made a BS with FBT
technology in our experimental realization, and found that the
coupling ratio is 0.5 in the 1550 nm wavelength, while the 1470 nm
and 1290 nm source have the coupling ratio 0.986 and 0.003
respectively. Interestingly, we can apply the 1470 nm (1290 nm)
source to control the selection of the rectilinear basis (diagonal
basis) in Bob's side.
 Using this loophole, we present that Eve can
control Bob's measurement basis choice remotely at the cost of only
increasing quantum bit error rate(QBER) from $1.3\%$ to $1.4\%$.

The FBT BS is made by closing two or more bare optical fibers,
fusing them in a high temperature environment and drawing their two
ends at the same time, then a specific biconic tapered waveguide
structure can be formed in the heating area. The FBT BS can be used
as the splitter or the coupler, it has the feature of low insertion
loss, good directivity and low cost, so many of the commercial BS
products are made by this technology. However, coupling ratio of the
FBT BS is wavelength-dependent, and most types of the FBT BS work
only in a limited range of wavelength (limited bandwidth), where the
coupling ratio of the BS is defined as
$r=\frac{I_{port1}}{I_{port1}+I_{port2}}$, $I_{port1}$ ($I_{port2}$)
is output light intensity from BS's output port 1 (output port 2).
Typical coupling ratio at the center wavelength provides optimal
performance, but the coupling ratio varies periodically with
wavelength changes. We made a BS with FBT technology in our
experimental realization, and found that it has distinguishing
wavelength-dependent characteristic, detailed expression of this
property can be given in Fig. 1.

\begin{figure}[!h]\center
\resizebox{7cm}{!}{
\includegraphics{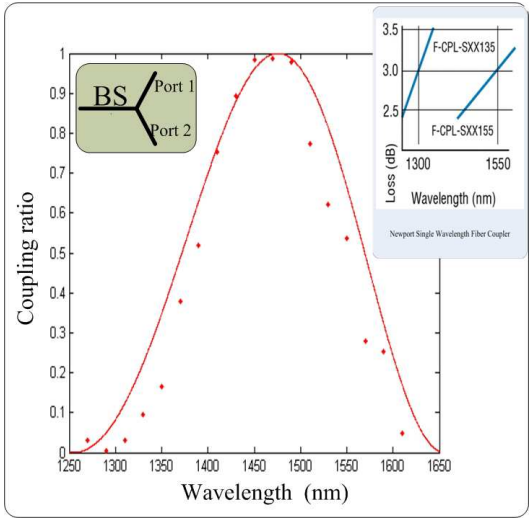}}
\caption{The relationship between wavelength of the source and the
coupling ratio, here the red dot is the practical experimental
result, the red line is the theoretical analysis result. The right
side is the single wave length fiber coupler made by the Newport
Corporation \cite{newport}, the coupling ratio of which are also
wavelength-dependent. }
\end{figure}

We analyze the relationship between wavelength $\lambda$ and
coupling ratio $r$ by using the coupling model given in Ref.
\cite{BS1,BS2}:
\begin{equation}
\begin{array}{lll}
r=F^2sin^{2}(\frac{Cw}{F}),
\end{array}
\end{equation}
where $F^2$ is the maximal power that is coupled,
$C\propto\lambda^{2.5}$ is the coupling coefficient, $w$ is the heat
source width. From Fig. 1, we can find that the realistic BS has the
perfect coupling ratio $0.5$ with 1550 nm laser diode (LD), in which
case the BS can be regarded as perfect QKD device. When we test it
with 1290nm LD and 1470nm LD, the coupling ratio changed to be 0.003
and 0.986, which means that the 1290 nm and 1470 nm LD will mainly
pass through BS's port 2 and prot 1 respectively. Thus the realistic
BS can not be regarded as perfect QKD device in case of wavelength
of the LD is not 1550 nm. Combining this imperfection with multi
wavelength sources, we show that Eve can acquire all secret key
information in Bob's side with very low cost \cite{lowe cost}.

The polarization based QKD system with passive state modulation can
be depicted precisely in Fig. 2.
\begin{figure}[!h]\center
\resizebox{9cm}{!}{
\includegraphics{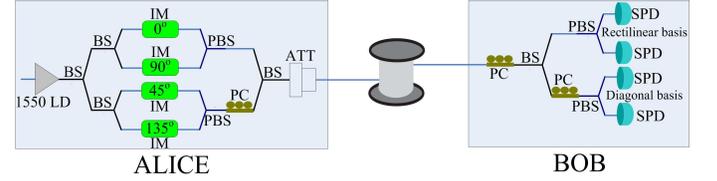}}
\caption{Schematic diagram of the polarization based QKD system,
where 1550 LD means Alice send the quantum state with the 1550 nm
source, BS is the beam splitter, PBS is the polarization beam
splitter, IM is the intensity modulator, PC is the polarization
controller, ATT is the attenuator, SPD is the single photon
detector.}
\end{figure}
After two cascaded BS with an additional intensity modulation, four
polarization states can be generated by 1550 nm LD. More precisely,
when Alice want to transmit the prepared quantum state, the positive
voltage will be added on the matched IM, and the negative voltage
will be added on the other IM respectively. Thus only the single
photon state modulated by the positive voltage can be transmitted
into the quantum channel. In the ideal polarization based QKD
experimental realization, one of the basic assumption is that the
photon state will pass through each output side with $50\%$
probability. Actually, this perfect BS in Bob's side can be regarded
as the random bases selector. Unfortunately, the coupling ratio of
the realistic FBT BS is wavelength-dependent as illustrated in the
previous section. Eve can adopt intercept-and-resend strategy to
attack practical polarization based QKD systems, where Eve's
detection setup in the quantum channel is the same as Bob's side.
Applying her state measurement result, Eve will send the remodulated
photon state to Bob. In this attacking protocol, the main difficult
for Eve is to find the appropriate LD with wavelengths $\lambda_1$
and $\lambda_2$, where $\lambda_1$ LD has the coupling ratio
$r_1>0.5$, $\lambda_2$ LD has the coupling ratio $r_2<0.5$. To
attack the practical QKD system, Eve will send the re-modulated
quantum state with $\lambda_1$ ($\lambda_2$) LD to Bob, if she can
get the detection result with the rectilinear basis $\{0
\textordmasculine,90\textordmasculine\}$ (diagonal basis $\{45
\textordmasculine,135\textordmasculine\}$).

We initially give the security analysis in the theoretical aspect
under the assumption that only the BS in QKD system is imperfect. By
considering intercept-and-resend strategy has been applied by Eve in
the quantum channel, the final QBER between Alice and Bob  can be
given by

\begin{equation}
\begin{array}{lll}
Err=\frac{1}{4}(\frac{1-r_1}{2-(r_1+r_2)}+\frac{r_2}{r_1+r_2}),
\end{array}
\end{equation}
this equation can be simply calculated with the probability tree of
the state transformation as illustrated in Fig. 3.
\begin{figure}[!h]\center
\resizebox{8cm}{!}{
\includegraphics{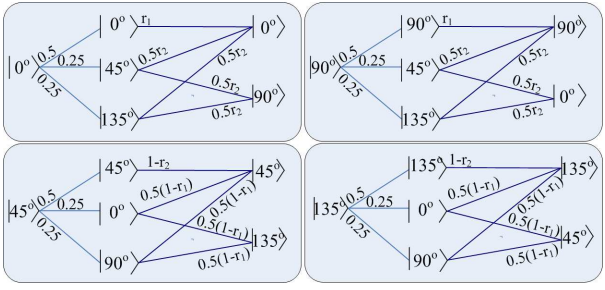}}
\caption{Probability tree of the state transformation. Alice sends
the modulated quantum sate to the quantum channel, Eve gets the
state detection result with probability $p_1\in\{0.25,0.25,0.5\}$ in
the middle stage, Bob saves his sate detection result after the
sifting protocol with probability $p_2\in
\{\frac{1}{2}r_1+\frac{1}{4}r_2,\frac{1}{4}r_2\}$ or
$\{\frac{1}{2}(1-r_2)+\frac{1}{4}(1-r_1),\frac{1}{4}(1-r_1)\}$ with
different measurement bases.}
\end{figure}
Utilizing Shor and Preskill's security analysis result with the
perfect QKD \cite{Shor}, Alice and Bob can distill the final secret
key if the QBER introduced by the eavesdropper is lower than $11\%$.
In case of the coupling ration and the wavelength have a strong
correlation ($r_1\rightarrow1, r_2\rightarrow0$), Eve can get full
secret key bit even if the error rate is lower \cite{Explanation}.
We note that no secret key can be established if the error rate is
lower than $Err$ between two legitimate parties. More interestingly,
even zero QBER can not generate any secret key with full wavelength
dependent BS ($r_1=1, r_2=0$).

By using the analyzed realistic BS in the previous section, detailed
setup of the attacking system can be illustrated in Fig. 4.
\begin{figure}[!h]\center
\resizebox{9cm}{!}{
\includegraphics{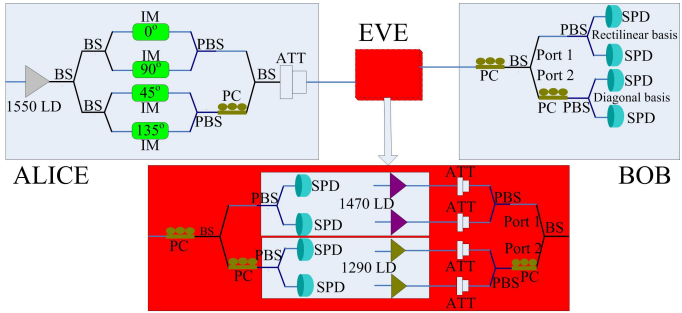}}
\caption{Attacking practical polarization based QKD system. The red
area is controlled by the eavesdropper Eve, who will utilize the
intercept-and-resend strategy by applying the wavelength-dependent
BS and multi-wavelength sources. }
\end{figure}
In this system, if Eve get the measurement result 0 (1) with the
rectilinear basis $\{0 \textordmasculine,90\textordmasculine\}$, she
will prepare the quantum state $|0 \textordmasculine\rangle$
($|90\textordmasculine\rangle$) again with the 1470 nm LD.
Conversely, if she can get the detection result 0 (1) with the
diagonal basis $\{45 \textordmasculine,135\textordmasculine\}$, she
will prepare the quantum state  $|45 \textordmasculine\rangle$
($|135\textordmasculine\rangle$) again with the 1290 nm laser diode,
where $|45 \textordmasculine\rangle=\frac{1}{\sqrt{2}}(|0
\textordmasculine\rangle+|90\textordmasculine\rangle),|135\textordmasculine\rangle=-\frac{1}{\sqrt{2}}(|0
\textordmasculine\rangle-|90\textordmasculine\rangle)$. We give a
simple example that Alice sends the quantum sate $|0
\textordmasculine \rangle $, and Eve gets the detection result $|0
\textordmasculine\rangle $ in the rectilinear basis $\{0
\textordmasculine,90\textordmasculine\}$ with probability $50\%$,
then she will send the remodulated 1470 nm laser to the receiver
Bob, since the 1470 nm laser can mainly pass through the port 1 of
the BS in Bob's side, Bob can detect $|0 \textordmasculine \rangle $
in the rectilinear basis  with $98.6\%$ success probability. If Eve
get the detection result in the diagonal basis  $\{45
\textordmasculine,135\textordmasculine\}$ with the probability
$50\%$, then she will send the remodulated 1290 nm laser to the
receiver Bob, since the 1290 nm laser can mainly pass trough the
port 2 of the BS in Bob's side, Bob can get the detection result in
the diagonal basis with $99.7\%$ success probability. Note that the
detection efficiency of the practical SPD is also
wavelength-dependent, we verified that the id 200 SPD \cite{ID} has
detection efficiency $12.1\%$, $10.7\%$ and $5.0\%$ by considering
wavelengths of the source are 1550 nm, 1470 nm and 1290 nm
respectively. To solve this problem, we will add different
attenuations after the 1470 nm LD and 1290 nm LD, thus Bob can get
the similar detection result in case of with and without the
eavesdropper.

Following the attacking model given in the previous section, we give
the photon count result in Bob's side by considering two cases:
without the eavesdropper and with the eavesdropper respectively. In
the first case, Alice randomly send the polarization state $\{|0
\textordmasculine \rangle,|90\textordmasculine
\rangle,|45\textordmasculine \rangle,|135\textordmasculine\rangle\}$
to the quantum channel. Considering the practical single photon
detection efficiency is $12.1\%$ in 1550 nm case, we can get about
$1\%$ effective detection result in case of the quantum channel has
10.79 dB attenuation. In our practical experimental realization
without the eavesdropper, Alice send $10^6$ prepared quantum states,
then Bob get about $10^4$ detection results, which can be
illustrated precisely in Fig. 5.

\begin{figure}[!h]\center
\resizebox{10cm}{!}{
\includegraphics{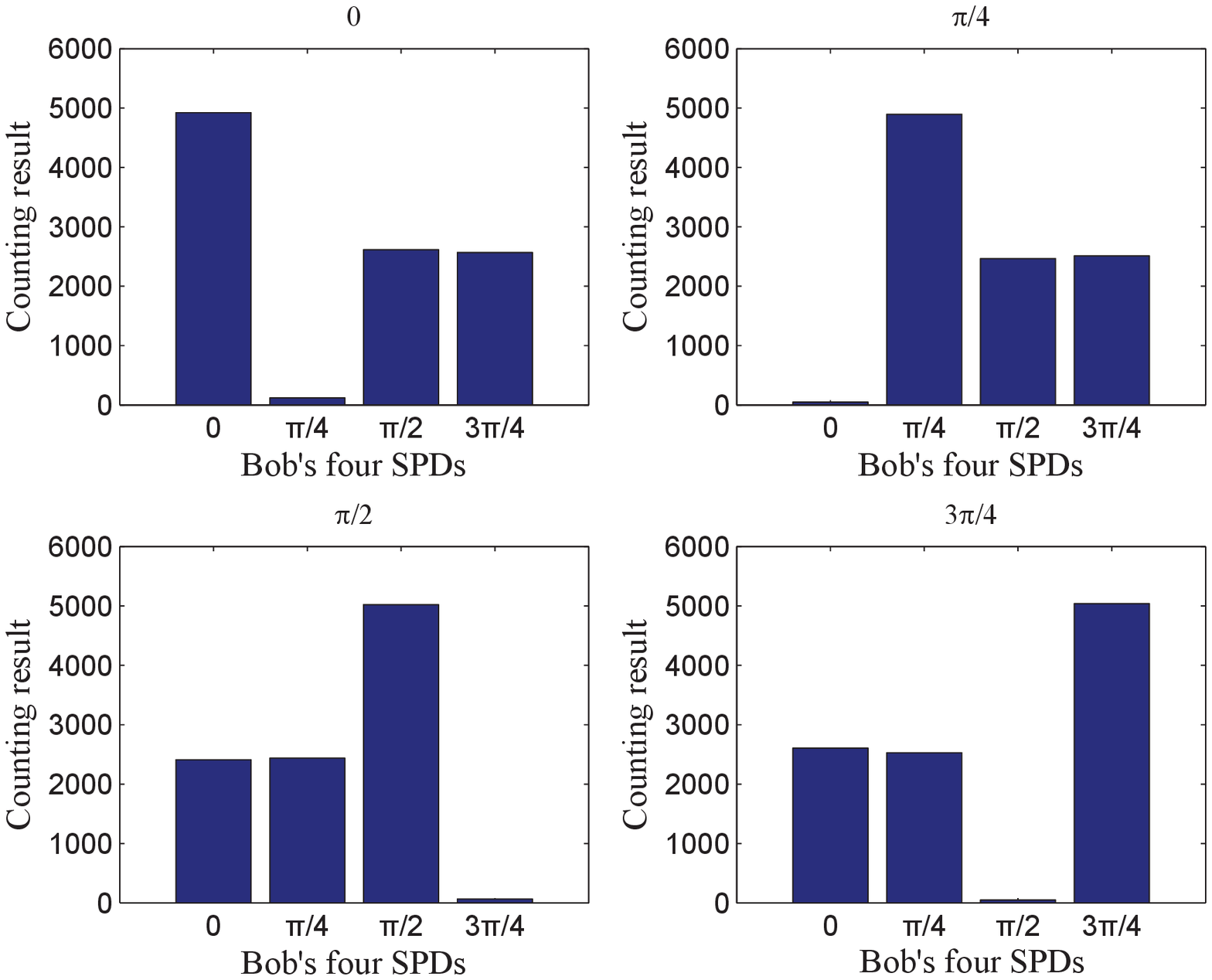}}
\caption{The detection result in Bob's side without the eavesdropper
Eve. Alice sends four quantum states $\{|0 \textordmasculine
\rangle,|90\textordmasculine \rangle,|45\textordmasculine
\rangle,|135\textordmasculine\rangle\}$ to the quantum channel with
1550 nm LD, In Bob's side, he can get the correct detection result
if the matched basis has been chosen. He will get the detection
result with $50\%$ error rate if the unmatched basis has been chosen
correspondingly.}
\end{figure}

In case of with the eavesdropper, Eve will apply the similar
detection setup, but the channel attenuation between Alice and Eve
is 0 dB, the reason for which is that Eve can utilize the lossless
channel instead of the standard optical fiber.  We give the
detection result in Bob's side by considering Eve send $5\times10^3$
modulated quantum states (each pulse has two photons on average) to
the quantum channel, the channel loss between Eve and Bob is 3.3 dB
and 0 dB in 1470 nm and 1290 nm cases respectively, then Bob can get
about $5\times10^3$ effective detection results, detailed detection
results can be illustrated in Fig. 6.
\begin{figure}[!h]\center
\resizebox{10cm}{!}{
\includegraphics{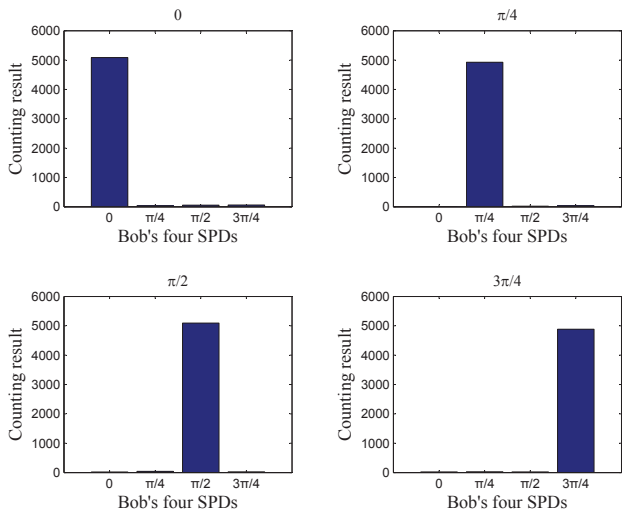}}
\caption{The detection result in Bob's side with the eavesdropper.
Eve sends quantum states $\{|0 \textordmasculine
\rangle,|90\textordmasculine \rangle\}$ with 1470 nm LD, sends
quantum states $\{|45\textordmasculine
\rangle,|135\textordmasculine\rangle\}$ with 1290 nm LD to the
quantum channel. In Bob's side, he can only get the detection result
in the rectilinear basis when the 1470 nm laser has been detected.
Similarly, he can only get the detection result in the diagonal
basis when the 1290 nm laser has been detected.}
\end{figure}
Comparing different types of detection results, we find that Eve can
remotely control Bob's basis selection only by changing LD's
wavelengths.

Based on this strategy, polarization based QKD system has been
attacked in our practical experimental realization. Bob gets the
similar detection result in case of with and without the
eavesdropper. Similarly, the QBER between Alice and Bob only
increases from $1.3\%$ to $1.4\%$, thus Eve can get almost all of
the secret key information without being discovered.

In conclusion, we propose a new type of strategy to attack the
practical polarization based QKD system by using the
wavelength-dependent BS and multi-wavelength sources. The
eavesdropper Eve can control Bob's measurement basis with $100\%$
success probability without reducing the receiver's expected
detection rate or significantly increasing the bit error rate. Our
result demonstrate that all practical devices require security
inspection for avoiding side channel attacks in practical QKD
experimental realizations. We note that this attacking protocol can
not be avoided even if the wavelength filter was applied in Bob's
side, since Eve only need increase the intensity of the light to
attack Bob's detection setup. Meanwhile, we should also note that
this attacking protocol can be avoided effectively by applying the
actively modulated phase encoding QKD systems \cite{Ph1, Ph2, Ph3}.

H.-W. Li thank X.-B. Z for helpful discussions, we thank N. Jain for
helpful comments and bringing Ref. \cite{Jain} to our attention, we
thank the anonymous referees, they had put a real great effort into
reviewing this article, and they had provided lots of useful
feedback that helped to improve the presentation of this article.
 This work was supported by the
National Basic Research Program of China (Grants No. 2011CBA00200
and No. 2011CB921200), National Natural Science Foundation of China
(Grant NO. 60921091), National High Technology Research and
Development Program of China (863 program) (Grant No. 2009AA01A349),
and China Post doctoral Science Foundation (Grant No.
20100480695).\\To whom correspondence should be addressed,
Email:\\$^a$kooky@mail.ustc.edu.cn,\\ $^b$yinzheqi@mail.ustc.edu.cn,\\ $^c$zfhan@ustc.edu.cn.\\
$^*$These authors contributed equally to this work.


\begin{thebibliography}{10}
\bibitem{lowe cost} In principle, the eavesdropper Eve is assumed to have unlimited computing and storage power when proving unconditional security of the ideal QKD protocol.
Our attacking model show that Eve only requires two practical
different wavelength sources, which implies that the strategy lowers
the cost but works more efficiently.
\bibitem{Explanation} Considering the strong correlation, the QBER between Alice and Bob
 is $Err\rightarrow0$. Bob gets the rectilinear basis detection result
 with probability $\frac{1}{2}(r_1+r_2)\rightarrow\frac{1}{2}$,
 gets the diagonal basis detection result with probability
 $1-\frac{1}{2}(r_1+r_2)\rightarrow\frac{1}{2}$. Thus Eve's operation without increasing unbalanced detection
 and QBER.
\end{thebibliography}
\end{document}